\documentclass[11pt,letterpaper]{aastex63}
\usepackage{amssymb,amsmath,mathrsfs,color}
\usepackage{bm}
\usepackage{hyperref}
\usepackage{multirow}
\usepackage{color}
\usepackage{array}
\usepackage{dcolumn}

\hypersetup{pdftitle={},
pdfauthor={Dustin Lang},pdfsubject={Astronomical image processing}}

\newcommand{\arxiv}[1]{\href{http://arxiv.org/abs/#1}{arXiv:#1}}

\newcommand{\foreign}[1]{\emph{#1}}
\newcommand{\etal}{\foreign{et\,al.}}

\newcommand{\figref}[1]{\figurename~\ref{#1}}
\newcommand{\Figref}[1]{\figref{#1}}
\newcommand{\appref}[1]{Appendix \ref{#1}}
\newcommand{\eqnref}[1]{Equation \ref{#1}}
\newcommand{\niceurl}[1]{\href{#1}{\textsl{#1}}}
\newcommand{\project}[1]{\textsl{#1}}
\newcommand{\conv}{\otimes}
\newcommand{\sersic}{S\'ersic}

\newcommand{\transpose}[1]{#1^{T}}
\newcommand{\CD}{C\!D}
\newcommand{\pixelfx}{\nu}
\newcommand{\pixelfy}{\omega}
\newcommand{\pixelfreqs}{$(\nu,\, \omega)$}

\definecolor{gray}{rgb}{0.5,0.5,0.5}


\begin{document}

\title{A hybrid Fourier--Real Gaussian Mixture method for fast galaxy--PSF convolution}
\author{Dustin Lang}
\affiliation{%
  Perimeter Institute for Theoretical Physics,
  31 Caroline Street North, Waterloo, ON N2L 2Y5, Canada}
\affiliation{%
  Department of Physics \& Astronomy,
  University of Waterloo,
  200 University Avenue West, Waterloo, ON N2L 3G1, Canada}
\date{Dec 31, 2020}
\shorttitle{Fast Galaxy--PSF Convolution}
\shortauthors{Lang}

\begin{abstract}
  I present a method for the fast convolution of a model galaxy
  profile by a point-spread function (PSF) model represented as a pixel grid.
  The method relies upon three
  observations: First, most simple radial galaxy profiles of common interest
  (deVaucouleurs, exponential, \sersic) can be approximated as
  mixtures of Gaussians.  Second, the Fourier transform of a Gaussian
  is a Gaussian, thus the Fourier transform of a mixture-of-Gausssian
  approximation of a galaxy can be directly evaluated as a mixture of Gaussians
  in Fourier space.
  Third, if a mixture component would result in Fourier-space aliasing, that
  component can be evaluated in real space.
  For mixture components to be evaluated in Fourier space, we
  use the FFT for the PSF model, direct evaluation of the Fourier
  transform for the galaxy, and the inverse-FFT to return the result to
  pixel space.
  For mixture components to be evaluated in real space---which only happens when the mixture
  components is much larger than the PSF---we use a simple Gaussian approximation of
  the PSF, perform the convolution analytically, and evaluate in real pixel space.
  The method is fast and exact (to the extent that the mixture-of-Gaussians
  approximation of the galaxy profile is exact) as long as the pixelized PSF
  model is well sampled.
  This Fourier method can be seen as a way of applying a perfect low-pass filter to
  the (typically strongly undersampled) galaxy profile before convolution by
  the PSF, at exactly the Nyquist frequency of the PSF pixel model grid.  In
  this way, it avoids the computational expense of a traditional super-resolution
  approach.
  This method allows
  the efficient use of pixelized PSF models (ie, a PSF represented as
  a grid of pixel values) in galaxy forward model-fitting approaches such as
  \project{the Tractor}.
\end{abstract}

\section{Introduction}

A number of codes exist to render images of galaxies as they would
appear when observed with a (real or imagined) telescope and camera,
and given observing conditions (sky brightness, atmosphere).  These
include \project{GalSim} \citep{galsim}, \project{Ufig} \citep{ufig},
\project{phosim} \citep{phosim}, \project{galfit} \citep{galfit}, 
\project{ngmix} \citep{ngmix},
\project{the Tractor} (Lang \etal, in prep), and several others.

%\project{MegaMorph} \citep{megamorph},

In codes that simulate at the pixel level rather than the photon
level, it is necessary to convolve a galaxy's appearance ``above the
atmosphere'' (at high resolution and before any effects from the
atmosphere, optics, or detector) by the point-spread function to
produce the image as it would appear on the CCD.  This step often
dominates the computation time required to render a galaxy.  Since
model galaxy profiles such as the deVaucouleurs profile are very
``peaky'', typical galaxy images ``above the atmosphere'' are
undersampled by the native resolution of the CCD.  Thus it is not
possible to render na\"ively the above-the-atmosphere galaxy at native
pixel scale and then convolve by the PSF (also represented at native
pixel scale), because undersampling will lead to aliasing.
Codes typically attempt to render the images at higher resolution than the
CCD pixels and then bin down to the native CCD resolution; this
significantly increases the computational cost, and may still not
achieve exactly correct results.

As an illustration of this problem, consider the limit of a tiny,
point-like galaxy.  If rendered ``above the atmosphere'' at the native
image pixel scale, nearly all its light will appear in one pixel,
regardless of its position within that pixel.  If we generate a series
of images, scanning the galaxy across different subpixel positions in
the image, the galaxy's pixelized light will jump from one pixel to
the next in steps.  When convolved by the PSF, the galaxy will look
like a point source in the image, but rather than moving smoothly will
jump from one pixel to the next.  If rendered at twice the native pixel
scale and binned, it will jump in half-pixel steps, and if rendered at
four times the native pixel resolution, will jump in quarter-pixel
steps.  This occurs because the ``above the atmosphere'' image is
undersampled; a point source has power at all frequencies, and
rendering it at the native pixel scale causes the high-frequency power
to alias into the band, causing errors in the rendering.  As example of this
is shown in \figref{fig:aliasing}.

Some of the codes mentioned above, and in particular \project{GalSim},
provide the option to render PSF-convolved galaxies in Fourier space.
To avoid the aliasing discussed above, it is necessary to evaluate the
Fourier transform of the galaxy profile directly in Fourier space.
For exponential galaxy profiles, the \project{GalSim} authors give an
analytic expression.  For deVaucouleurs and \sersic\ profiles,
however, a look-up table must be used, and they caution that a
separate look-up table must be pre-computed for each \sersic\ index,
which can dominate the computational cost.  Finally, due to the
periodic nature of the inverse Fourier transform, spatial aliasing can
occur if the galaxy profile is rendered into an image that is too
small (as illustrated in \figref{fig:wrap}).  If the \project{GalSim}
user requests such a rendering, it produces the rendering on a larger
image (possibly at considerable extra cost) and returns the un-aliased
sub-image.  The hybrid Fourier/real method presented here provides an
inexpensive alternative.

% FIXME -- and the method for approximating galaxies with general
% Sersic index allows the Sersic index to be fit cheaply.

In \project{the Tractor}, we have previously taken the approach of
using mixture-of-Gaussian approximations\footnote{
  A \emph{mixture of Gaussians} is a weighted sum of $N$ Gaussian
  components, $\sum_{i=1}^N a_i \, \mathcal{N}(\mu_i, \Sigma_i)$
  with $\sum_i a_i = 1$.}
for \emph{both} the PSF and
the model galaxy profiles.  The galaxy profile approximations are
presented in \cite{moggalaxy}, and we use direct chi-squared
minimization to fit mixture-of-Gaussian models to pixelized PSF
models.  With the PSF and galaxy represented as mixtures of Gaussians,
convolution is analytic and results in a larger mixture of Gaussians
(the product of the two mixture sizes), which can then be directly
evaluated in pixel space.  This approach has the distinct limitation
that the PSF must be approximated as a mixture of Gaussians; in
practice, it takes many Gaussians to achieve a good approximation, and
this makes the computation expensive.

In this work, I consider pixelized PSF models, such as those produced
by \project{PsfEx} \citep{psfex}.  Since these models are created from
real images, they are naturally \emph{pixel-convolved} models; that
is, they already include the effect of being sampled by square pixels.
%and we do not need to think about pixels as ``little boxes''; the PSF
%model is a sampling of an underlying smooth surface.

The galaxy models considered here are simple elliptical isophotes with
surface brightness defined, as a functional of radius $r$, by
\begin{equation}
  I(r) \propto \exp(- k(n) \, r^{1/n})
\end{equation}
where we can fix $n=1$ (exponential profile), $n=4$ (deVaucouleurs
profile), or fit for $n$ (\sersic\ profile).  It is also common to fit
for a linear combination of $n=1$ and $n=4$ profiles (``bulge-disk
decomposition'' or ``composite galaxy profile'').  The term $k(n)$ is
a scalar, usually defined for each $n$ so that half the light falls
within $r \le 1$. The elliptical shape can be represented in terms of
effective radius, axis ratio, and position angle, or a variety of
other ellipse representations.

\begin{figure}
  \begin{center}
    \begin{tabular}{@{}c@{\hspace{1ex}}c@{\hspace{1ex}}c@{}}
      % Pixel Space & Fourier Space (Analytic) & Fourier Space \\
      %
      %\cline{1-1} \cline{2-2} \cline{3-3}
      %
      Step 1: PSF
      &
      &
      FFT(PSF) \\
      %
      % psf img
      \includegraphics[height=0.22\textwidth]{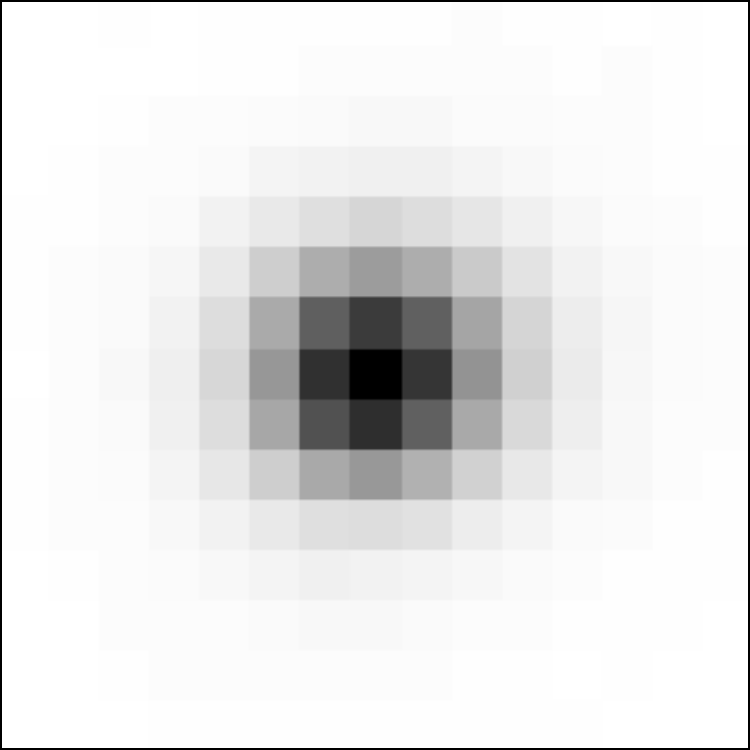}%
      & 
      \makebox[0em][c]{
        \raisebox{0.11\textwidth}{%
          \hspace{2em}$\xrightarrow{\displaystyle\textrm{\hspace{1em}
              Step 2: FFT \hspace{1em}}}$
        }%
      }
      &
      % psf FT
      %\multicolumn{1}{r}{%
      \makebox[0.22\textwidth][r]{%
        \includegraphics[height=0.22\textwidth]{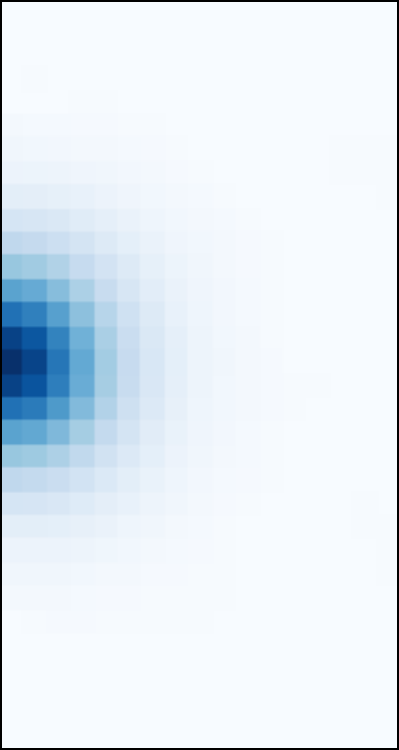}%
      }
      \\
      Step 3: Galaxy MoG
      &
      Step 5: FT(MoG)
      &
      Step 6: FT(Galaxy) \\
      %
      % gal img
      \includegraphics[height=0.22\textwidth]{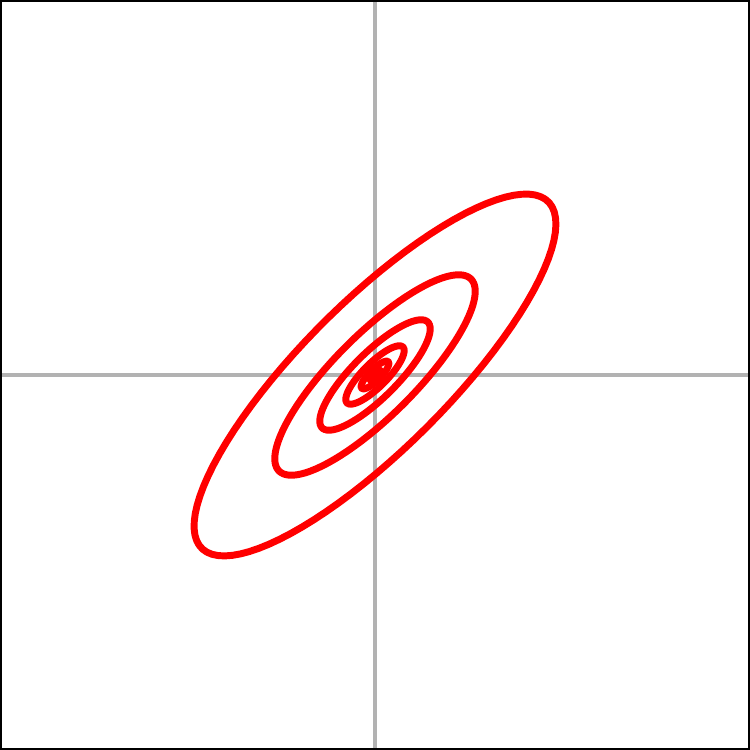}%
      &
      % gal analytic FT MoG
      \makebox[0.22\textwidth][r]{%
        \raisebox{0.11\textwidth}{$\longrightarrow$}%
        \hspace{1em}%
        \includegraphics[height=0.22\textwidth]{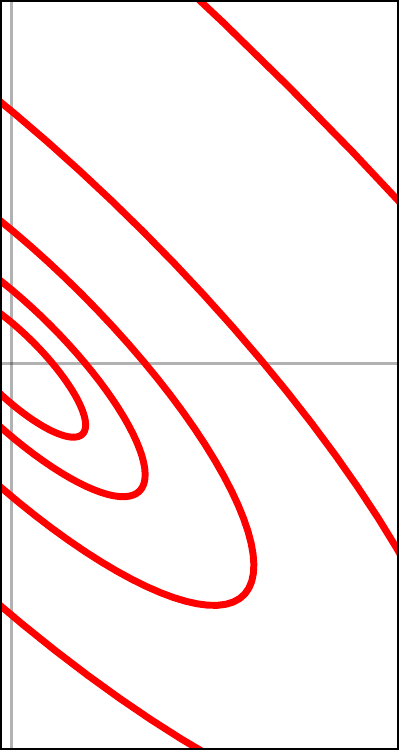}%
      }%
      \raisebox{0.11\textwidth}{\makebox[0ex][l]{%
          \hspace{1em}$\longrightarrow$}}
      \
      &
      % gal FT
      \makebox[0.22\textwidth][r]{%
        \includegraphics[height=0.22\textwidth]{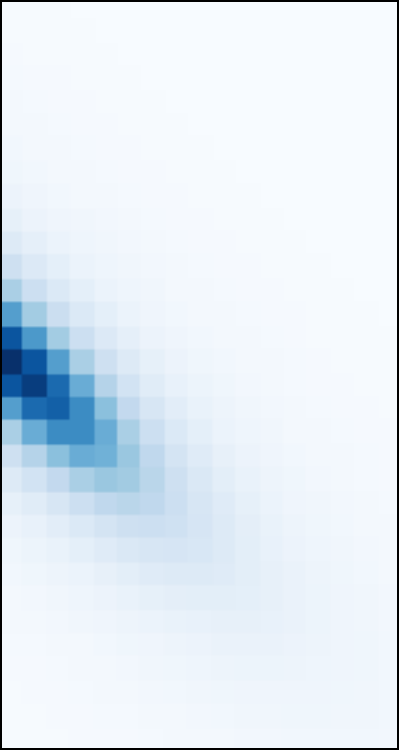}%
      }
      \\
      PSF $\conv$ Galaxy
      &
      &
      FT(PSF $\conv$ Galaxy) \\
      %
      % psf * gal img
      \includegraphics[height=0.22\textwidth]{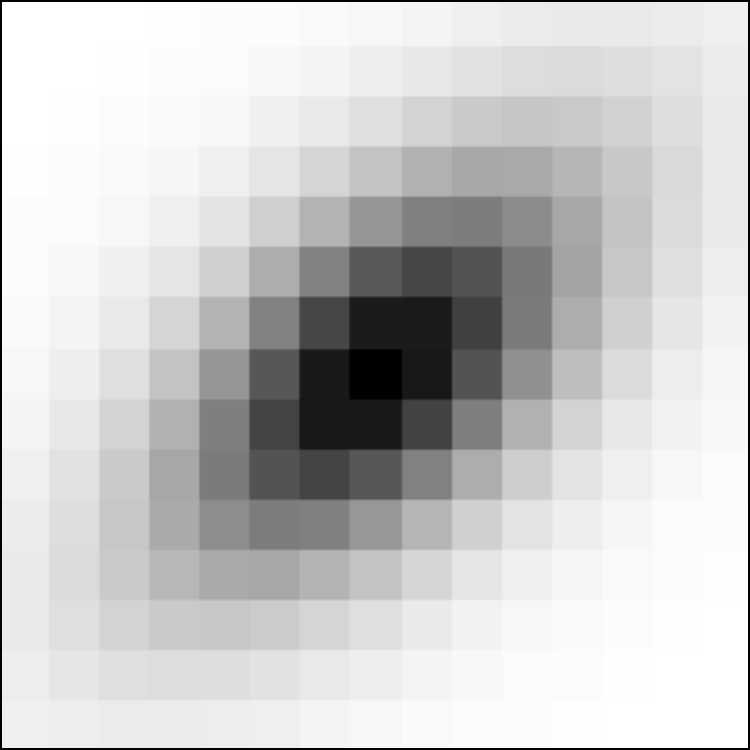}%
      &
      \makebox[0em][c]{
        \raisebox{0.11\textwidth}{%
          \hspace{2em}$\xleftarrow{\displaystyle%
            \textrm{\hspace{1em} Step 7: FFT$^{-1}$ \hspace{1em}}}$
        }%
      }
      &
      % psf * gal FT
      \makebox[0.22\textwidth][r]{%
        \includegraphics[height=0.22\textwidth]{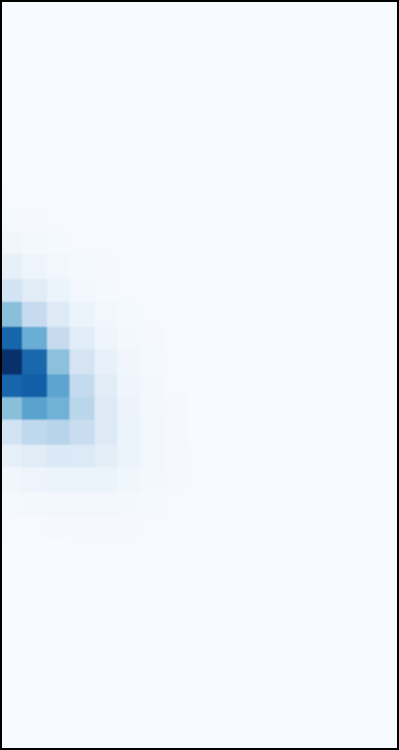}%
      }
    \end{tabular}
  \end{center}
  \caption{\label{fig:example}%
    Illustration of the method, for the case when all mixture components are to be
    evaluated in Fourier space.
    \textbf{Top-left}: pixelized PSF model (zoom-in of central region).
    \textbf{Top-right}: 
    FFT of the pixelized PSF model.  Since the PSF model is
    purely real, we show only half the symmetric Fourier transform. The FFT is
    complex; we show here the absolute value.
    By the assumption that the PSF is well-sampled, all Fourier components outside
    this frequency box are zero.
    \textbf{Middle-left}: Pixel-space galaxy profile,
    Mixture-of-Gaussians (MoG) approximation.  The profile would be
    undersampled if evaluated in pixel space; in this method, the galaxy model
    is never evaluated in pixel space.
    \textbf{Middle}: The mixture-of-Gaussians galaxy approximation in
    Fourier space.
    \textbf{Middle-right}: The discrete Fourier-space representation of the
    galaxy profile, evaluated directly in Fourier space by computing the value
    of each of the Gaussian components.
    \textbf{Bottom-right}: Product of the FFT of the PSF and the
    Fourier transform of the galaxy model.
    \textbf{Bottom-left}: Taking the inverse FFT of the bottom-right panel,
    we achieve the result: the galaxy model convolved by the PSF, in pixel space.
  }
\end{figure}

\section{The method}

The method presented here builds directly upon our mixture-of-Gaussians
approximation of galaxy profiles, with the realization that the
Fourier transform of a Gaussian is also a Gaussian.  This allows us to
evaluate the Fourier transform of a model galaxy profile \emph{directly} in Fourier
space, without ever having to render the ``above the atmosphere''
(unconvolved) galaxy profile in pixel space.
% (which is problematic due to undersampling).
We can then multiply this Fourier-space
representation of the galaxy by the Fourier transform of the pixelized
PSF model, then inverse Fourier transform back to pixel space to get
the PSF-convolved galaxy profile.
As long as the pixelized PSF model is well sampled, this method produces
exact answers for each Gaussian in the mixture.

We will elaborate on each step of the method below.  In summary:
\begin{enumerate}
\item Choose the desired output image size and embed the pixelized PSF
  model in a zero-padded image of this size.
\item Compute the Fourier Transform of the padded PSF model via the
  Fast Fourier Transform (FFT),
  recording the frequencies \pixelfreqs\ of the FFT terms.
\item Transform the mixture-of-Gaussians approximation of the model
  galaxy through an ellipticity matrix and the local image World
  Coordinate System (WCS) matrix to get pixel-space Gaussians
  (optionally including sub-pixel shifts).
\item Determine which (if any) components of the mixture-of-Gaussians
  galaxy model are large enough to produce significant spatial
  aliasing in the chosen output image size; omit these from the
  Fourier-space evaluation and evaluate in real space instead.
\item Convert the galaxy mixture-of-Gaussians from pixel space to Fourier space,
  analytically.
\item Evaluate the Fourier-space Gaussians on the frequency grid \pixelfreqs.
\item Multiply the PSF and galaxy Fourier transforms, then inverse-FFT
  back to pixel space.
\item Add any galaxy Gaussian mixture terms evaluated in real space,
  then, optionally, use Lanczos interpolation to perform sub-pixel
  shifting, if not done in Fourier space.
\end{enumerate}
%We will elaborate on each step below.

The process is illustrated in Figure \ref{fig:example}.  Example code
is available alongside the paper source,\footnote{ Available at
  \niceurl{https://github.com/dstndstn/fft-gal-convolution}.}  and the
method is also implemented in \project{the Tractor} code.\footnote{%
  Available at
  \niceurl{https://github.com/dstndstn/tractor}.}

Step 1, choosing the desired output image size, depends on the size of
the PSF model as well as the galaxy.  If the chosen size is too small,
wrap-around spatial aliasing will occur since the FFT is periodic.
This is mitigated by Step 4 (evaluating galaxy mixture-of-Gaussian
terms that will result in aliasing in real space rather than Fourier
space), but it remains necessary to choose a large enough output image
to capture a sufficient fraction of the light from the galaxy model.
In practice, one would likely choose the PSF size plus some factor of
the galaxy half-light radius to capture a sufficient fraction of the
flux, or, taking the brightness of the galaxy into account, a
sufficient surface brightness accuracy.

In Step 2, we compute the FFT of the PSF.  If the PSF is constant
across the image, this FFT need only be computed once per image.
Alternatively, some PSF modelling tools, including \project{PsfEx}
\citep{psfex}, produce \emph{eigen-PSF} models.  That is, the PSF
model varies spatially across the image, and the variation is
represented as polynomial variations in the amplitudes of a set of
pixelized PSF components.  This is convenient, since the FFT of each
eigen-component can be computed once in preprocessing and the FFT at a
given position in the image can be computed by weighting the component
FFTs using the PSF model's polynomial spatial variation terms.

In Step 3, we transform the mixture-of-Gaussians approximation of the
galaxy profiles from their representation in units of galaxy effective
radius into pixel units.  This includes an ellipse transformation
(scaling by effective radius, shearing by galaxy axis ratio, rotation
by galaxy position angle), plus a transformation into pixel
coordinates based on a locally-linear approximation of the image's
World Coordinate System.
The result is a pixel-space mixture-of-Gaussians representation of the
galaxy.

Specifically, in \project{the Tractor}, we prefer to represent the
shape of a galaxy by the effective radius $r_e$ and two ellipticity
components $e_1$ and $e_2$ as are typically used in weak lensing.  We
can represent this as a matrix that transforms from celestial
coordinates into units of effective radius (the coordinates in which
the galaxy profiles are written).  This is most easily done by
computing the position angle $\theta$ and axis ratio $a$,
\begin{align}
\theta & = \frac{1}{2} \arctan\!2(e_2, e_1) \\
e & = \sqrt{e_1^2 + e_2^2} \\
\beta & = \frac{1 - e}{1 + e} \label{eq:a}
\end{align}
leading to the transformation matrix
\begin{align}
E &= r_e \begin{bmatrix}
\textcolor{white}{-}\beta \cos{\theta} & \sin{\theta} \\
-\beta \sin{\theta} & \cos{\theta} \\
\end{bmatrix} \quad .
\end{align}
If we represent the local astrometric world coordinate system by an
affine transformation that takes celestial coordinates to pixel
coordinates, then we can multiply the two transformation matrices to
convert from pixels to galaxy profile coordinates.  The FITS WCS
standard \citep{wcs2} defines a transformation matrix $\CD$ that takes
pixels to intermediate world coordinates.  The combined affine transformation
matrix $A$ is then
\begin{align}
A & = (\CD)^{-1} E
\end{align}
so that a covariance matrix in galaxy effective radius coordinates can
be transformed into pixel space by computing
\begin{align}
C &= A \, V \transpose{A}
\label{eq:vpix}
\end{align}
where the galaxy profile's variances are isotropic; $V = v \bm{I}$.
When we write out this matrix in full, we find that the $\cos{\theta}$
and $\sin{\theta}$ terms always appear in pairs, so we can avoid
computing these trigonometric functions explicitly in the above.  See
\appref{app:transform}.

\begin{figure}[htb]
\begin{center}
\begin{tabular}{@{}ccc@{}}
  $128 \times 128$-pixel FFT &
  $32 \times 32$ FFT &
  $32 \times 32$ Hybrid \\
  \includegraphics[height=0.19\textwidth]{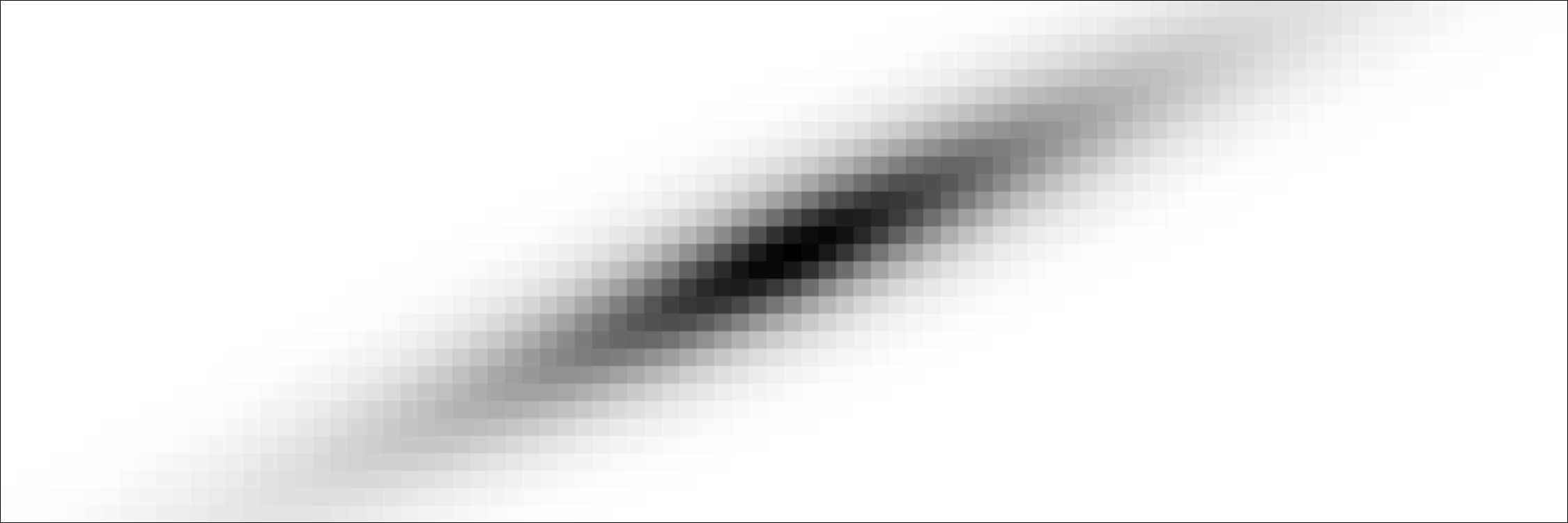} &
  \includegraphics[height=0.19\textwidth]{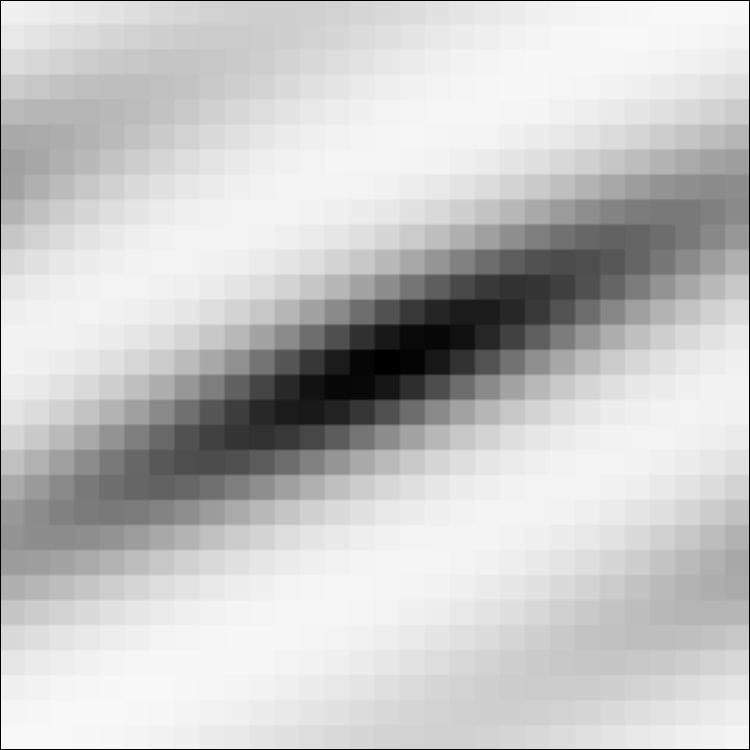} &
  \includegraphics[height=0.19\textwidth]{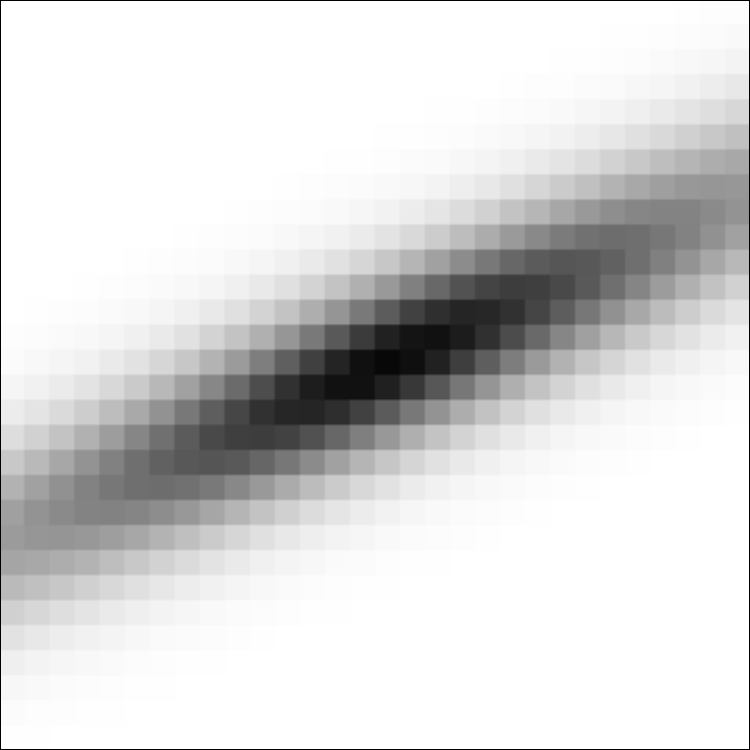} \\
\end{tabular}
\end{center}
\caption{\label{fig:wrap}%
  Spatial (wrap-around) aliasing of the PSF-convolved galaxy profile
  can occur if the FFT is
  computed on a pixel grid too small to contain the galaxy.
  Our hybrid approach---switching from Fourier space to real space
  for galaxy Gaussian mixture components that would alias---eliminates
  this problem.
  \textbf{Left:} Result (cropped) when galaxy is evaluated in a 128-pixel square grid in Fourier space.
  \textbf{Middle:} Result when evaluating the FFT in a 32-pixel square grid shows wrap-around
  aliasing.
  \textbf{Right:} Result when using our hybrid method shows negligible aliasing.
}
\end{figure}

In Step 4, we determine which components of the mixture-of-Gaussians
representation of the galaxy profile will result in significant
spatial aliasing in the FFT.  In \project{the Tractor}, we compute the
sum of the diagonal elements of the pixel-space covariance matrix $C$,
take the square root and divide this by half the output image size;
this is roughly the number of `sigmas' from the center to the edge of
the image, for that Gaussian component.  For Gaussian mixture
components where this is greater than $3$, we evaluate in Fourier
space.  For components less than $4$, we evaluate in real space.  For
components between $3$ and $4$, we compute both real and Fourier
versions and ramp smoothly between them, so that the galaxy profile
has continuous first derivatives.  To evaluate in real space, we use a
single-Gaussian approximation of the PSF.  Since this approximate PSF
model is only used when the galaxy model is much larger than the PSF,
this approximation is reasonable.  See \figref{fig:wrap}.

Step 5 is the core of the method.  The Fourier transform of a Gaussian
is another Gaussian; our mixture-of-Gaussian representation of galaxy
profiles can be converted into a mixture of Gaussians in Fourier
space.
A single 2-dimensional Gaussian
centered at the origin and
with covariance $C$
in pixel space is defined by
\begin{equation}
g(\bm{x}) = \frac{1}{2 \pi \sqrt{\det(C)}}
\exp\left( -\frac{1}{2} \transpose{\bm{x}} C^{-1} \bm{x} \right)
\end{equation}
which becomes, writing out the pixel coordinate vector $\bm{x}$
as $\transpose{(x,y)}$
and symmetric covariance $C = \bigl(\begin{smallmatrix}
a&b \\ b&d
\end{smallmatrix} \bigr)$,
\begin{equation}
g(x, y) = \frac{1}{2 \pi \sqrt{a d - b^2}}
\exp \left(
-\frac{d x^2 - 2 b x y + a y^2}{2(a d - b^2)}
\right) \quad .
\end{equation}

We can perform sub-pixel shifts of the galaxy model either in Fourier
space, or by Lanczos convolution in pixel space.  If we choose to
shift in Fourier space, the Fourier Transform of the shifted Gaussian
$g(x - x_0, y - y_0)$ is defined as
\begin{equation}
G(\pixelfx,\, \pixelfy) =
\int\limits_{-\infty}^{\infty}
\int\limits_{-\infty}^{\infty}
g(x - x_0, y - y_0) e^{-2 \pi i (\pixelfx x + \pixelfy y)} \, \mathrm{d}x \, \mathrm{d}y
\quad ,
\end{equation}
and we use the shift theorem to move the centroid to $\mu = (x_0, y_0)$
in pixel space via a phase term in Fourier space;
\begin{equation}
G(\pixelfx,\, \pixelfy) =
e^{-2 \pi i (x_0 \pixelfx + y_0 \pixelfy)}
\int\limits_{-\infty}^{\infty}
\int\limits_{-\infty}^{\infty}
g(x, y) e^{-2 \pi i (\pixelfx x + \pixelfy y)} \, \mathrm{d}x \, \mathrm{d}y
\end{equation}
which, plugging in the covariance matrix $C$, works out to
\begin{equation}
G(\pixelfx,\, \pixelfy) =
e^{-2 \pi i (x_0 \pixelfx + y_0 \pixelfy)}
e^{-2 \pi^2 (a \pixelfx^2 + 2 b \pixelfx \pixelfy + d \pixelfy^2)}
\label{eq:onegaussian}
\end{equation}
which can easily be evaluated on a grid of frequencies \pixelfreqs.  This
is the expression we must evaluate to produce the Fourier transform of
a single Gaussian component of a galaxy model.  Recall that the values
$a$, $b$ and $d$ are simple manipulations of the galaxy radius and
ellipticity and the local astrometric transformation, while $x_0$ and
$y_0$ handle sub-pixel shifts.

For a mixture of Gaussians, we must evaluate $G$ for each component in
the mixture and compute their weighted sum.  In the case of a
mixture-of-Gaussians representation of affine-transformed galaxy
profiles, the centers $(x_0, y_0)$ are the same for each component in
the mixture; only the covariances,
$C_i = \bigl(\begin{smallmatrix}
a_i&b_i \\ b_i&d_i
\end{smallmatrix} \bigr)$,
differ.  The Fourier transform
of a mixture of Gaussians,
\begin{equation}
m(x,y) = \sum_i A_i \, g_i(x, y)
\end{equation}
is therefore
\begin{equation}
M(\pixelfx,\, \pixelfy) = e^{-2 \pi i (x_0 \pixelfx + y_0 \pixelfy)}
\sum_i A_i \,
e^{-2 \pi^2 (a_i \pixelfx^2 + 2 b_i \pixelfx \pixelfy + d_i \pixelfy^2)}
\quad .
\end{equation}

In practice, computing the phase shift in Fourier space requires many
$\sin$ and $\cos$ operations, so in \project{the Tractor} we instead
use Lanczos interpolation to perform sub-pixel shifting of the final
pixel-space model.  This also has the desirable effect that galaxy
models with tiny radius approach point-source models, to higher
precision than we find in practice when using Fourier phase shifts.

In Step 6 of our method, we directly evaluate these Gaussians in Fourier
space, on the same discrete frequency grid \pixelfreqs\ used for the FFT
of the PSF.  This has interesting properties.  By taking the Fourier
transform, we are effectively assuming that the PSF model is well
sampled, thus has zero power outside the frequency grid of the FFT.
The galaxy profile may have power outside that frequency range; this
is the reason that one cannot simply render the galaxy profile ``above
the atmosphere'' at the native image pixel scale and apply FFT
convolution na\"ively: In the limit, a point-like galaxy has power at
all frequencies.  But when we multiply the PSF and galaxy profile in
Fourier space, all that high-frequency power will be zeroed out by the
assumption of a well-sampled PSF model.  Effectively, by evaluating the
galaxy Fourier transform directly in the frequency space of the PSF
model, we are applying a perfect low-pass filter at exactly the
Nyquist limit.  By zeroing out the high-frequency power, we avoid the
aliasing that would otherwise result from Fourier transforming an
undersampled signal.

In Step 7, we multiply the PSF and galaxy Fourier transforms, and
apply the inverse FFT to return to pixel space.  Finally, in Step 8,
we add in the Gaussian mixture terms that were evaluated in real
space, and, if not done in Fourier space, use Lanczos interpolation to
perform a sub-pixel shift.

% Next, in \figref{fig:psfmog}, I show the results of using my method
% versus our previous approach of using a mixture-of-Gaussians
% approximation for the PSF.  We would expect to see the greatest
% differences for compact galaxies, and for PSFs that have asymmetries
% or broad wings that tend not to be fit well by small mixtures of
% Gaussians.
% 
% \begin{figure}
%   \begin{center}
%   \end{center}
%   \caption{\label{fig:psfmog}
%   }
% \end{figure}

\section{Discussion}

\begin{figure}[p!]
  \newcommand{\F}{$\mathcal{F}$}
  \begin{center}
    \begin{tabular}{@{}ccccc@{}}
      This method
      %& My galaxy FT
      &
      log(\F(This))
      %& Naive $-$ FSGM \\
      &
      \\
      \includegraphics[height=0.24\textwidth]{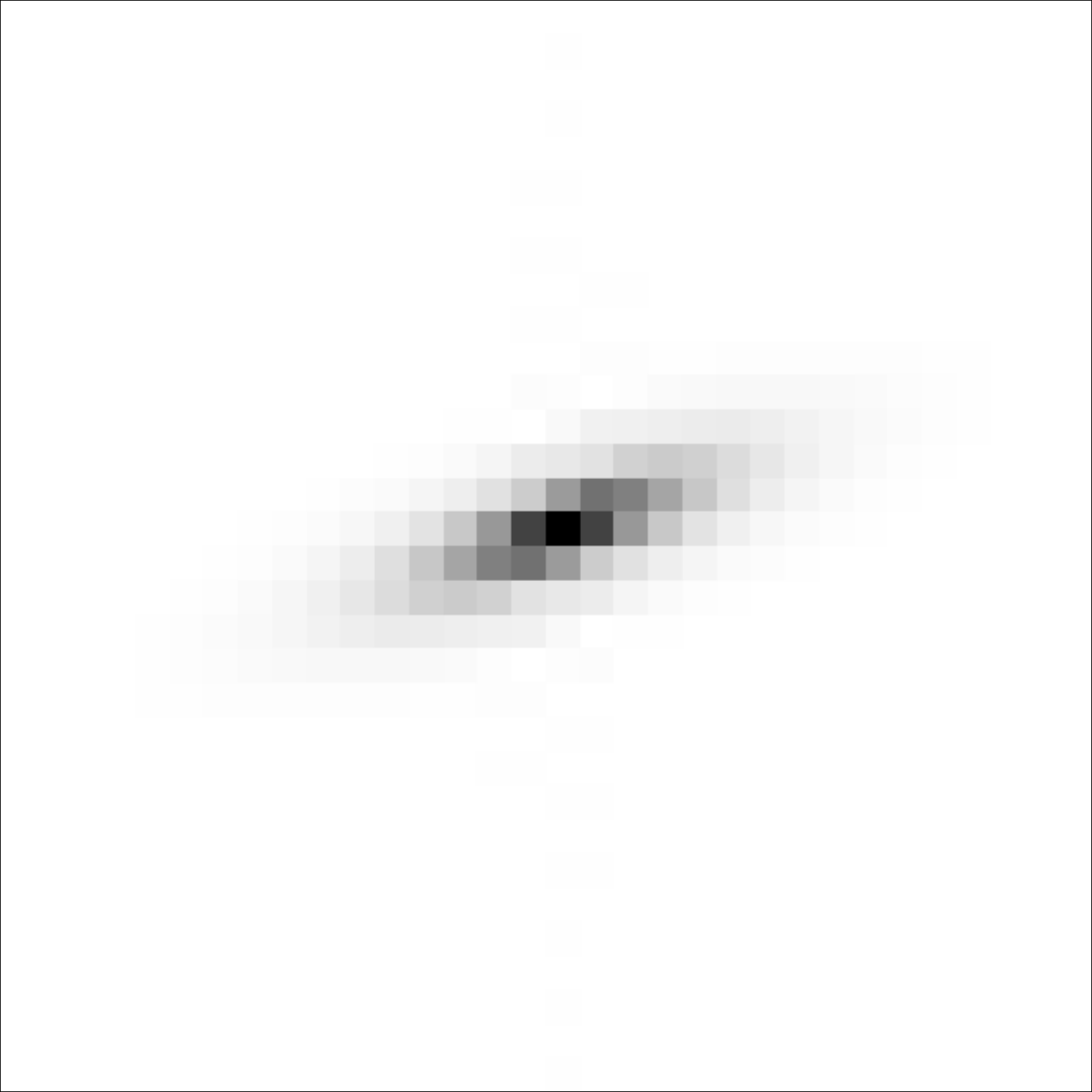}
      %&
      %$\leftarrow$
      %&
      %\includegraphics[height=0.24\textwidth]{lopass-mine-fourier}
      &
      \includegraphics[height=0.24\textwidth]{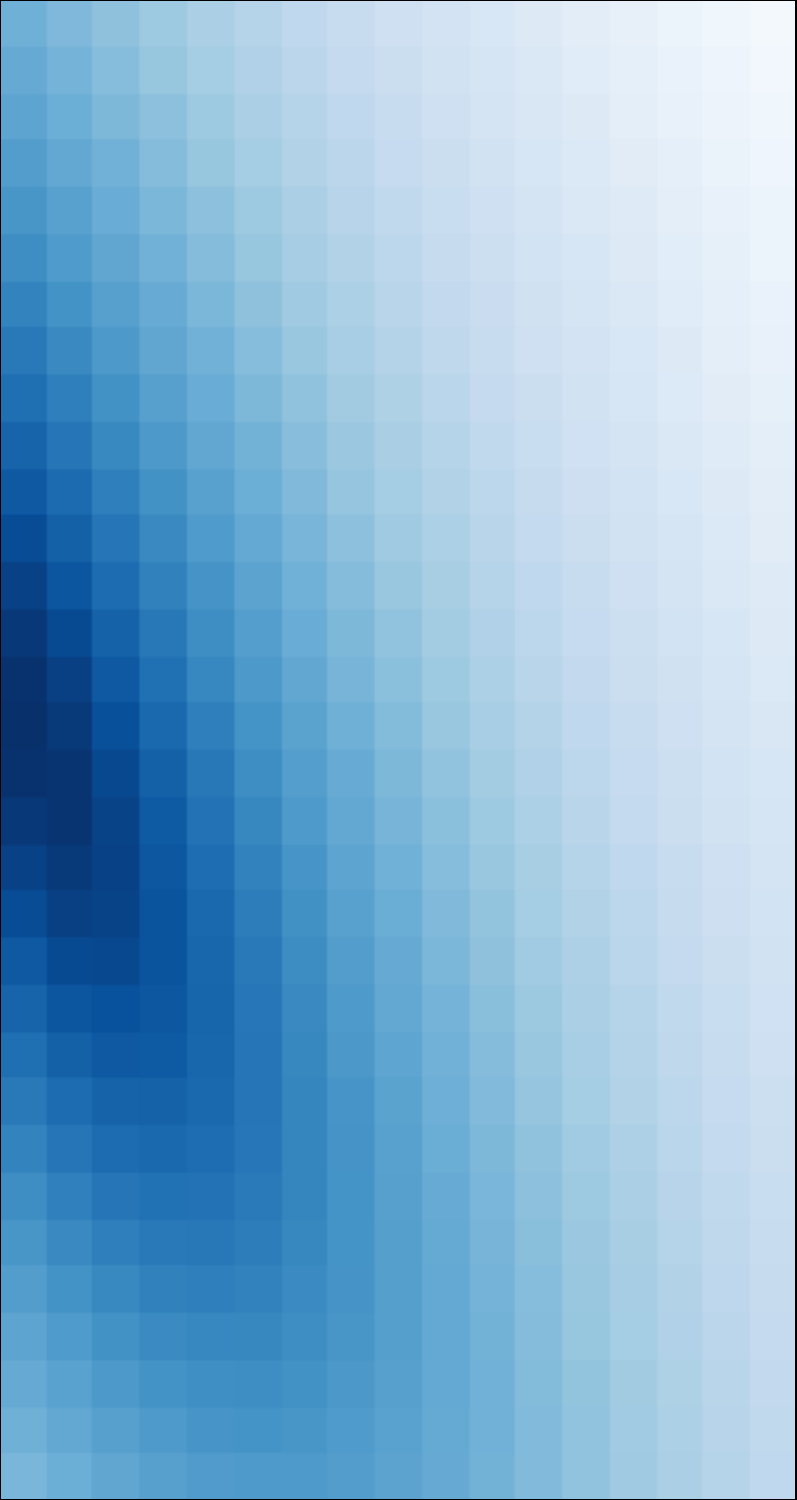}
      &
      \\
      Na\"ive
      %& Na\"ive galaxy FT
      & log(\F(Na\"ive))
      &
      \multicolumn{2}{c}{\F(Na\"ive $-$ This)}
      %&
      & Na\"ive $-$ This
      \\
      \includegraphics[height=0.24\textwidth]{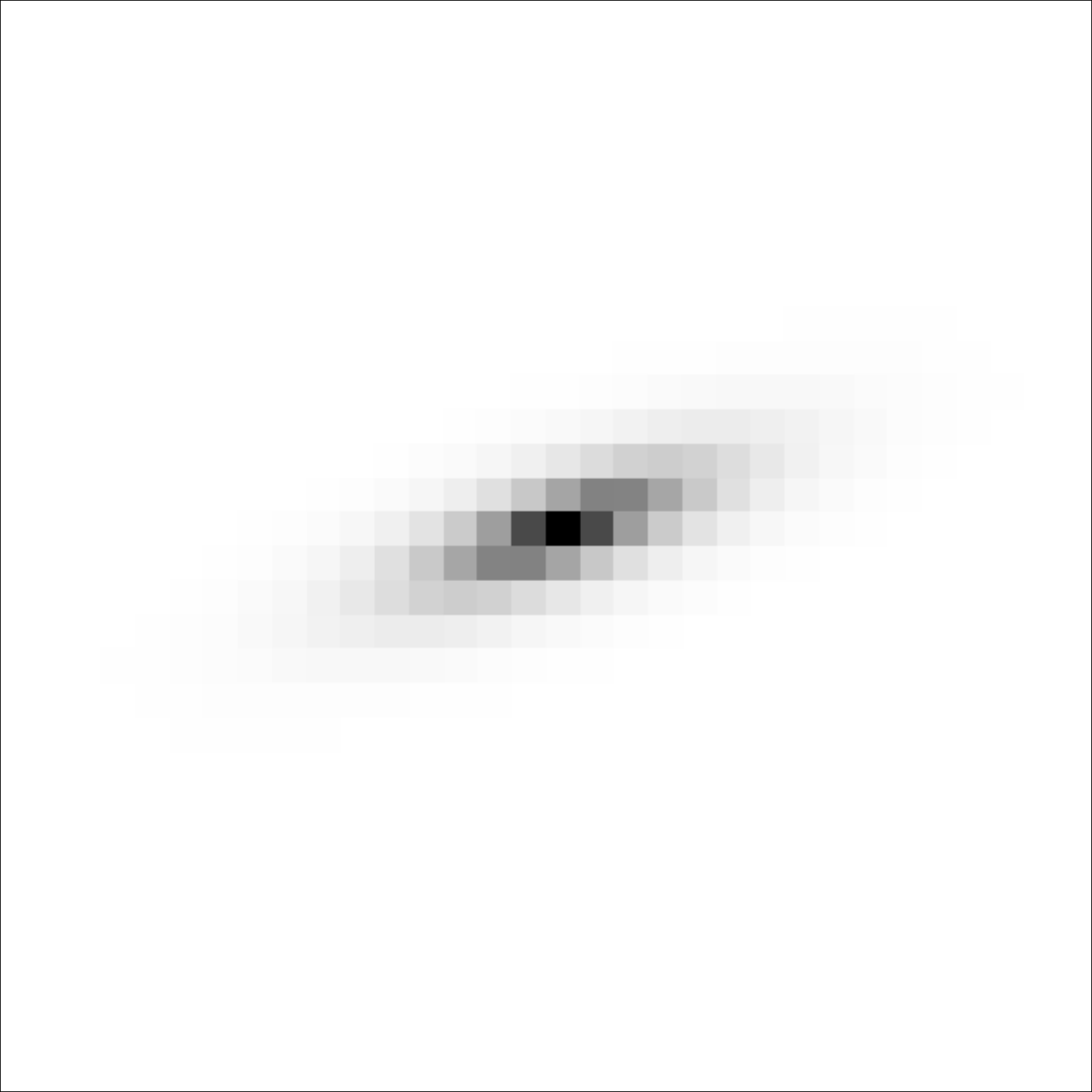}
      %&
      %$\rightarrow$
      %&
      %\includegraphics[height=0.24\textwidth]{lopass-naive-fourier}
      &
      \includegraphics[height=0.24\textwidth]{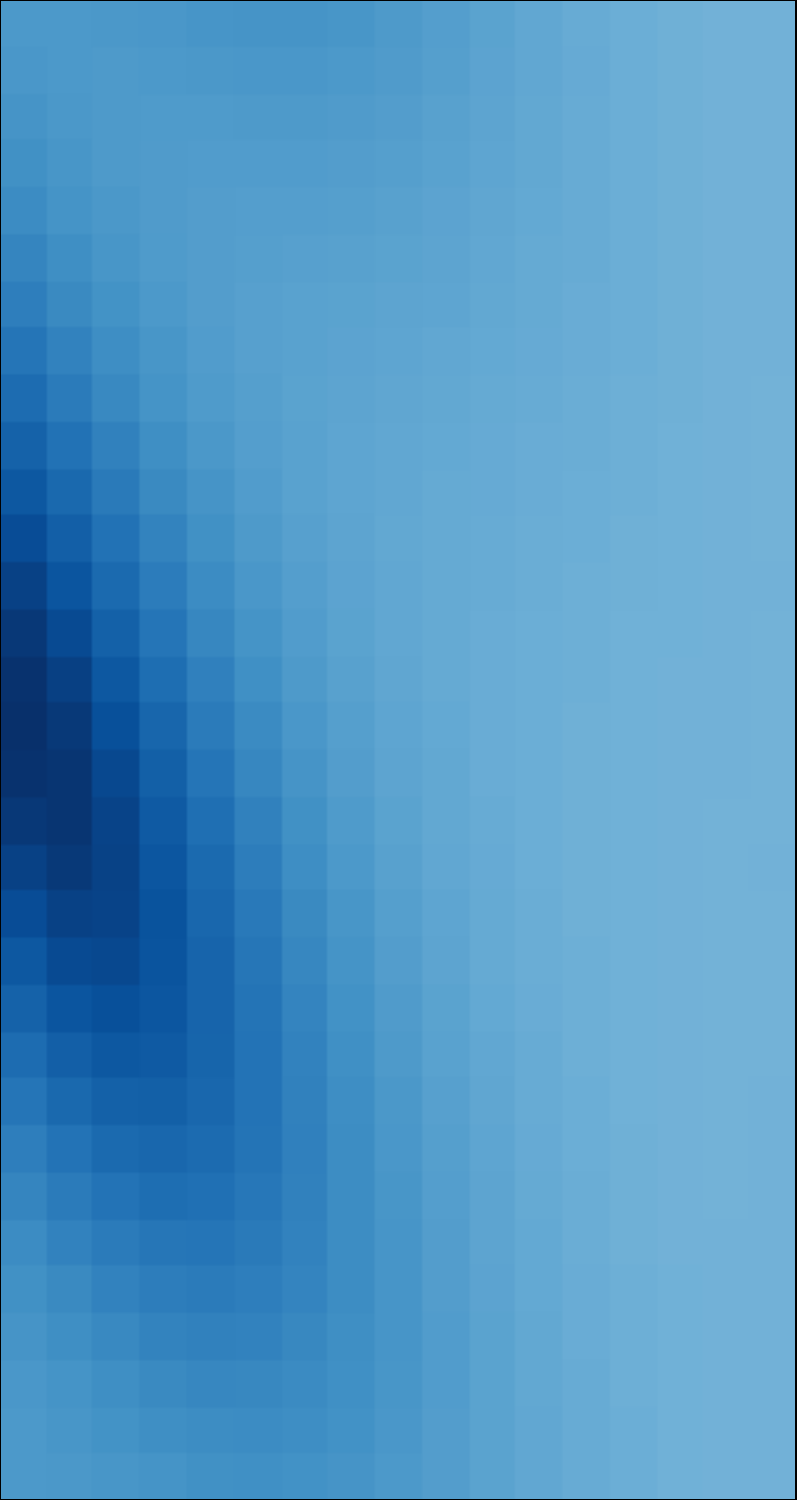}
      &
      \includegraphics[height=0.24\textwidth]{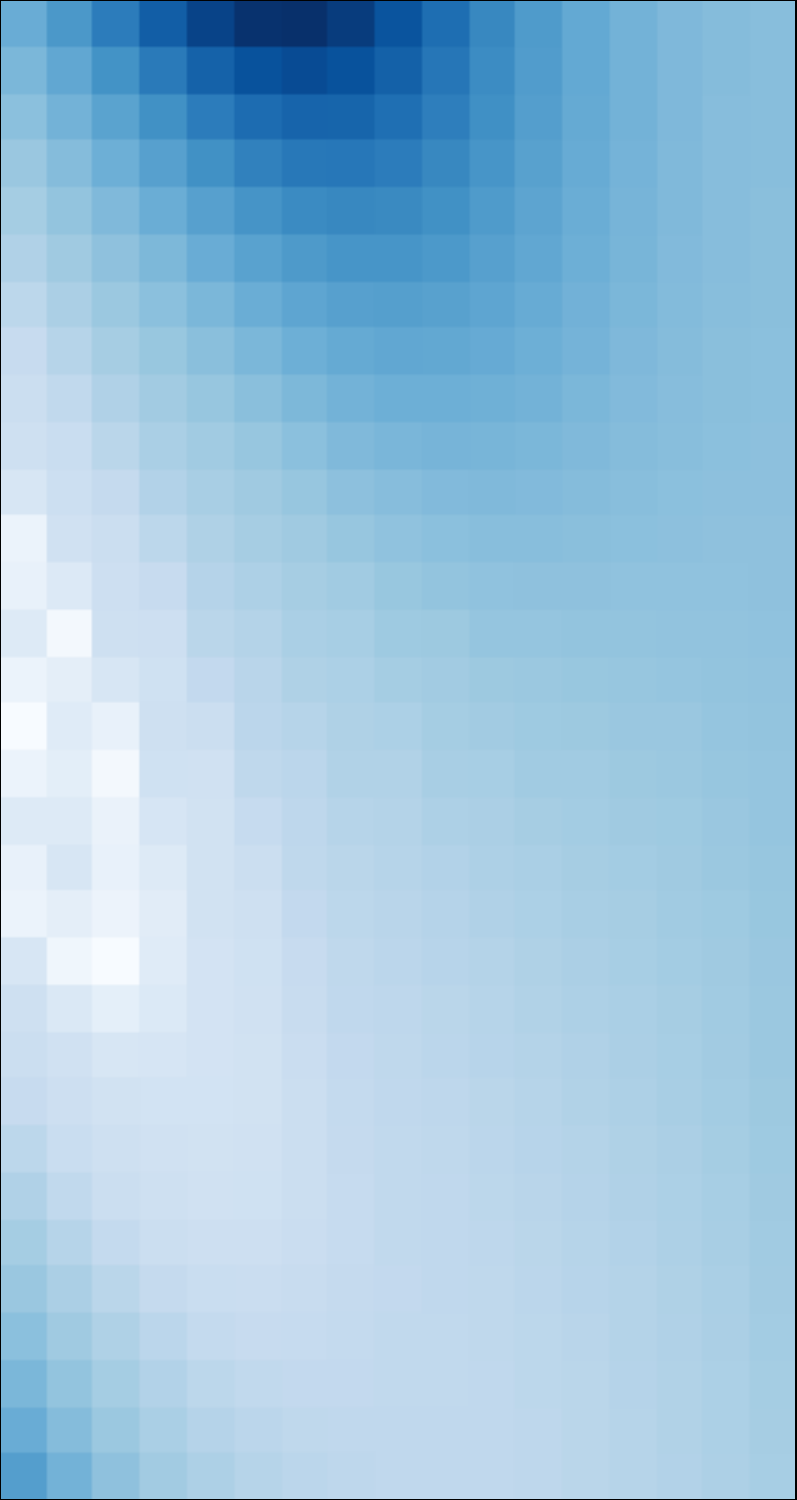}
      &
      \includegraphics[height=0.24\textwidth]{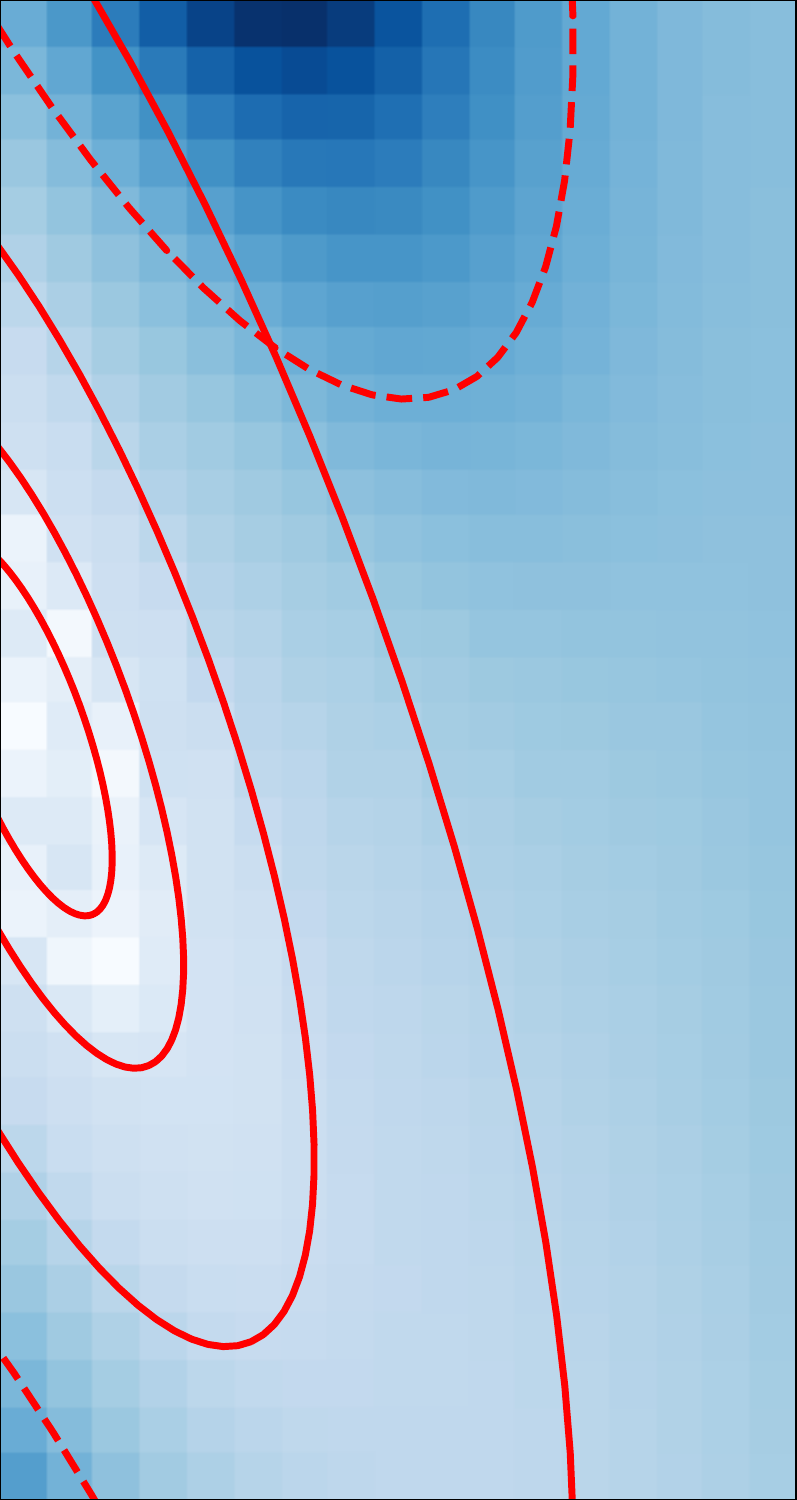}
      &
      \includegraphics[height=0.24\textwidth]{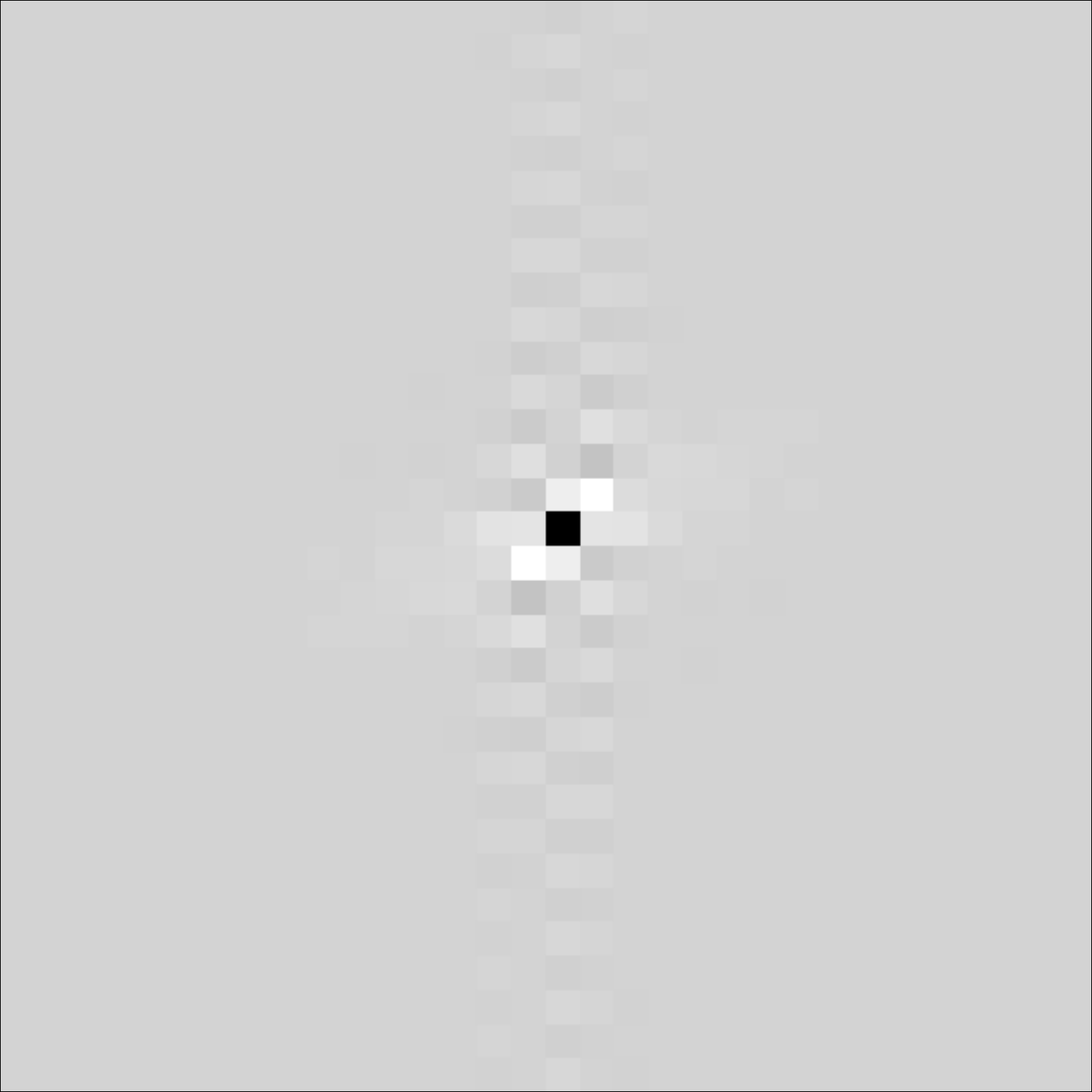}
      \\
      %%%
      Double-resolution
      &
      log(\F(Double))
      &
      \multicolumn{2}{c}{\F(Double $-$ This)}
      &
      Double $-$ This
      \\
      \includegraphics[height=0.24\textwidth]{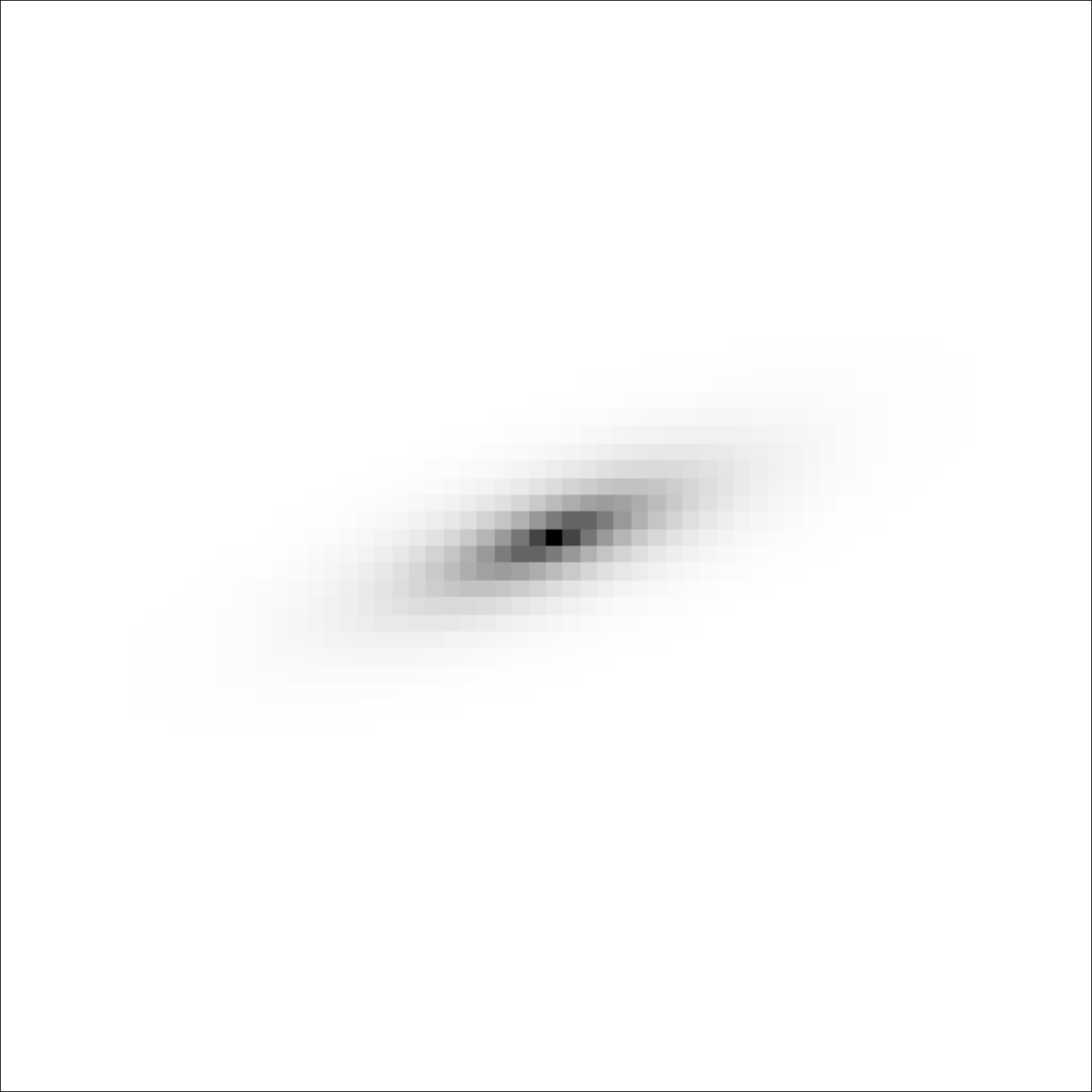}
      %&
      %\includegraphics[height=0.24\textwidth]{lopass-dclip-fourier}
      &
      \includegraphics[height=0.24\textwidth]{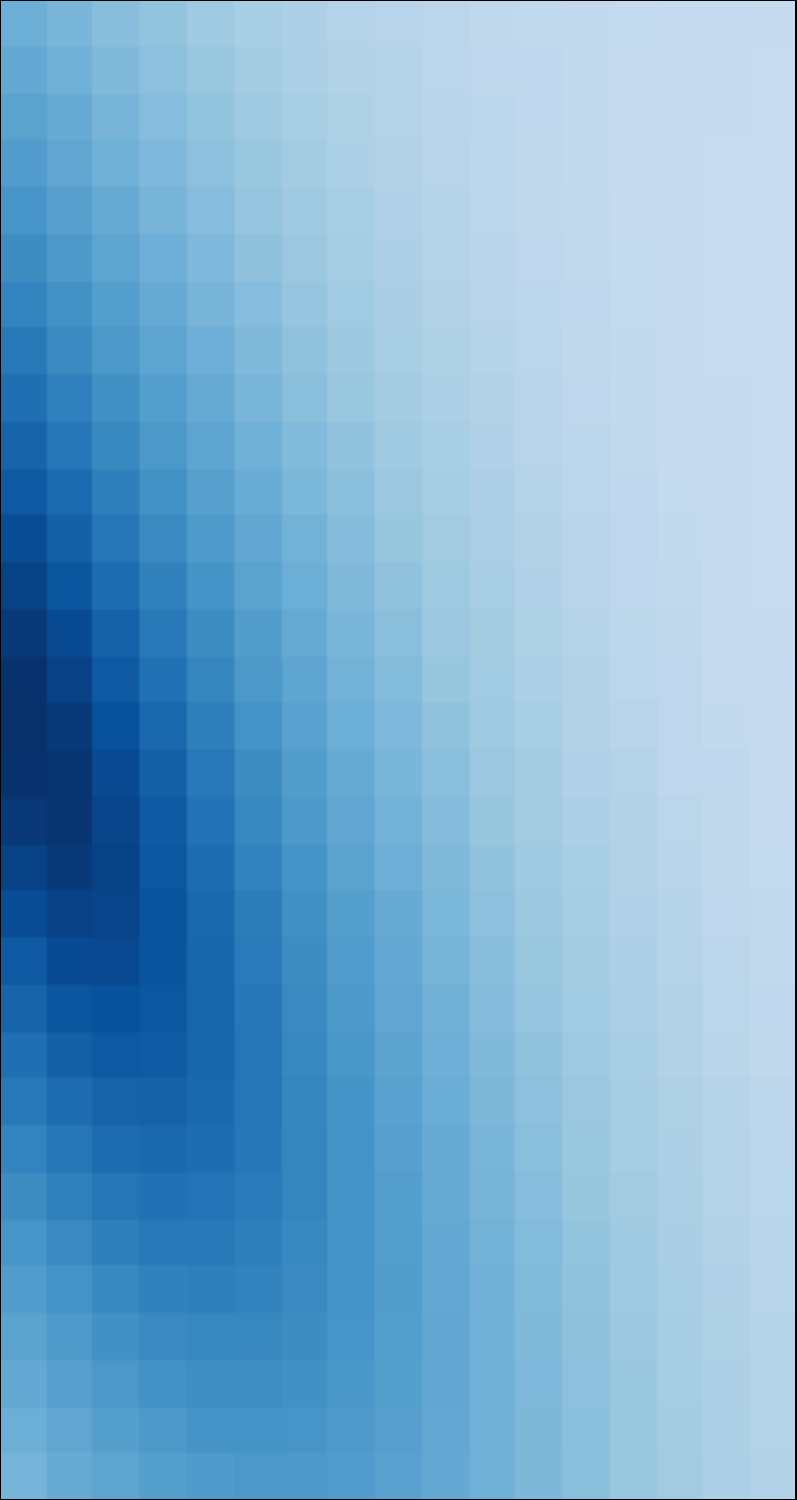}
      &
      \includegraphics[height=0.24\textwidth]{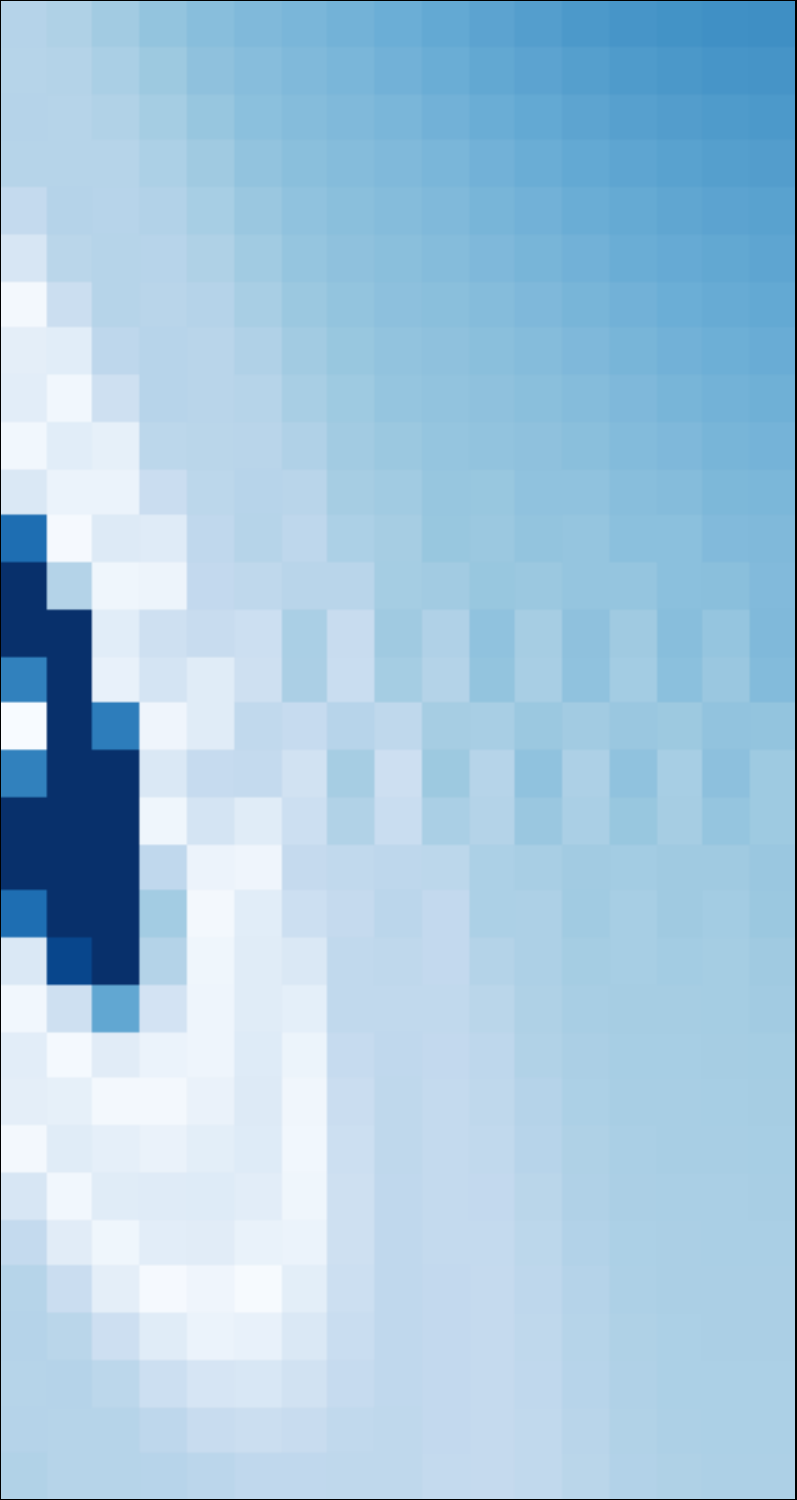}
      &
      \includegraphics[height=0.24\textwidth]{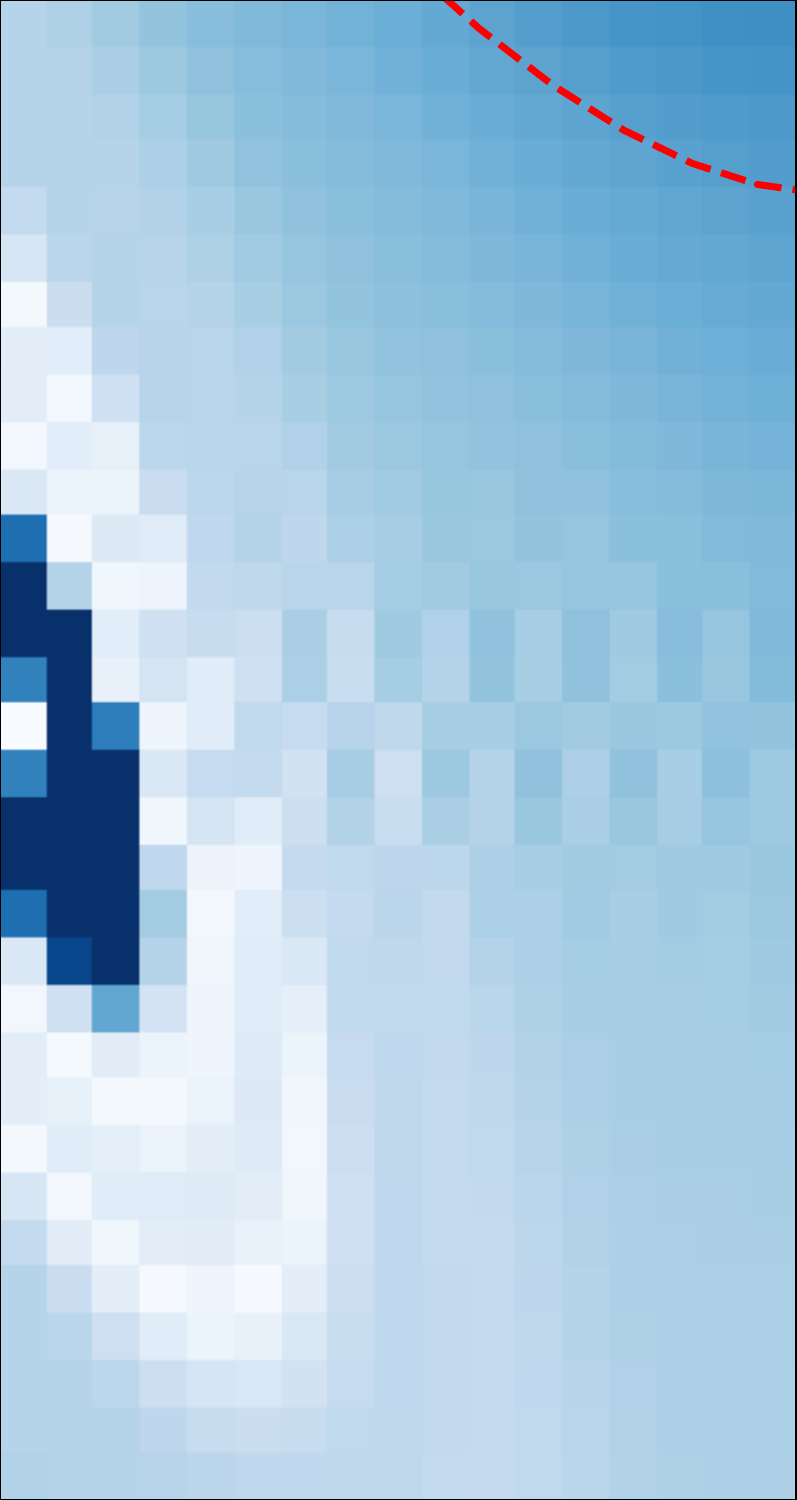}
      &
      \includegraphics[height=0.24\textwidth]{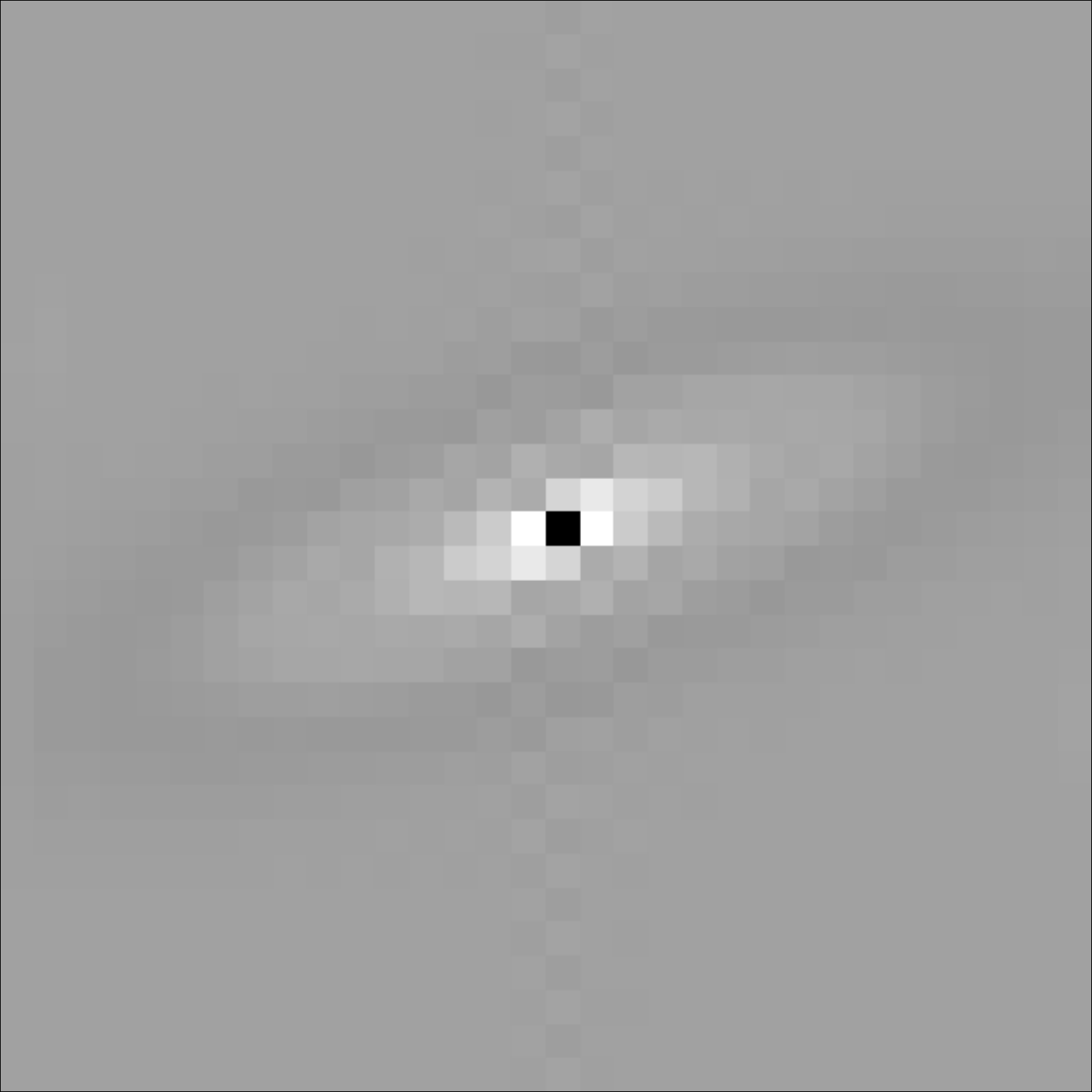}
      %\\
    \end{tabular}
    \caption{
      \label{fig:aliasing}
      An illustration of the high-frequency aliasing that results when
      a galaxy profile is rendered na\"ively at the typical native
      pixel scale of an optical image.  Shown here is a galaxy with an
      exponential radial profile and effective radius of 4 pixels.
      The core of this galaxy is strongly peaked, and is not well
      sampled.  The first row shows our results, the second row shows
      the under-sampled galaxy profile that results when the
      exponential profile function is evaluated at pixel centers, and
      the third row shows the still-undersampled profile that results
      when the profile is rendered at twice the resolution.
      The first column shows the pixel-space representation, and the
      second column shows Fourier space (log stretched).  The third
      column shows the difference from our results, and the fourth
      column marks the Gaussian mixture components in order to
      highlight where aliasing occurs.  A notable difference is the
      aliased lobes (shown with dashed lines), which appear because
      the Na\"ive galaxy model is undersampled, and the high-frequency
      power aliases into the low frequencies, causing high-frequency
      errors when the galaxy model is convolved by the PSF.  In the
      Double-resolution model, the aliased lobes are twice as far away
      from the origin, but still exist.  Another notable difference is
      that the central frequencies of the models differ.  This is
      expected, since our method is directly evaluating the Fourier
      transform of an approximation to the exponential galaxy profile.
      In the last column, we transform back to pixel space.  The
      high-frequency ``checkerboard'' strip results from ringing in
      the central pixels.
    }
  \end{center}
\end{figure}

This method for convolution of astronomical galaxy profiles with point
spread functions is both accurate and computationally efficient.  The
accuracy is demonstrated in \Figref{fig:aliasing}, surpassing the
accuracy of the significantly more expensive approach of using
sub-pixel bins in real space.  This method handles the real-world
complications of elliptical galaxy shapes and more complex galaxy
profiles such as \sersic\ profiles.  Code implementing the method is
publicly available in \emph{the
  Tractor}\footnote{\niceurl{https://github.com/dstndstn/tractor}}.

A lot of the computational cost of this method is in
evaluating the \emph{exp} function rather than in the FFT or
inverse-FFT, so fast \emph{exp} approximations\footnote{For example,
  \niceurl{https://github.com/herumi/fmath} as used in
  \project{ngmix}, \niceurl{https://github.com/esheldon/ngmix}.}
could provide significant speedups.
%
%http://homepage1.nifty.com/herumi/soft/fmath.html

% Although we have only discussed pixelized PSF models as the native
% pixel scale of the image, this method could also be used for
% situations where the images are undersampled but a higher-resolution
% pixelized PSF model is available (for example, Hubble Space
% Telescope).

The method is implemented in \project{the Tractor} code and has been
used at scale in the data reduction for the DESI Legacy Imaging Surveys
\citep{dey}, which include $20,000$ square degrees of imaging data from
three different cameras (the Dark Energy Camera \citep{decam},
the 90Prime camera \citep{90prime}, and the Mosaic3 camera \citep{mosaic3}),
measuring over two billion sources through forward modeling.

As mentioned earlier, when rendering \sersic\ or deVaucoulers galaxy
models in Fourier space, \project{GalSim} must pre-compute look-up
tables for the Fourier transforms, with a different table required for
each \sersic\ index.  The expense of this operation means that the
\project{GalSim} authors recommend using only discrete
\sersic\ indices.  We have recently found that the Gaussian mixture
components (amplitudes and variances) for \sersic\ galaxies vary
smoothly with respect to \sersic\ index, so we can fit for a discrete
set of indices and then interpolate for intermediate values.
This allows us to evaluate general \sersic\ models for the same
computational cost as exponential or deVaucoulers models.
Indeed, our latest Legacy Surveys release, Data Release 9, performs
these fits at scale and reports general \sersic\ fits for millions of
galaxies.

\acknowledgements

I thank Erin Sheldon (BNL),
David W.~Hogg (NYU), David J.~Schlegel (LBL), and John Moustakas (Siena College)
for helpful discussion and comments on this manuscript.

Research at Perimeter Institute is supported in part by the Government
of Canada through the Department of Innovation, Science and Economic
Development Canada and by the Province of Ontario through the Ministry
of Colleges and Universities.

\clearpage
\appendix

\section{Galaxy variances in pixel space}
\label{app:transform}

Writing out the matrix multiplication in \eqnref{eq:vpix} assuming an
isotropic variance $V = v I$, we get the pixel-space covariance matrix
\begin{align}
C = v f \begin{bmatrix}
C_{1,1} & C_{1,2} \\
C_{2,1} & C_{2,2}
\end{bmatrix}
\end{align}
where the matrix is symmetric so that $C_{1,2} = C_{2,1}$ and the terms are
\begin{align}
f & = \left( \frac{r_e}{t w - u v} \right)^2
\\
C_{1,1} & =
%
%\beta^2(w^2 c^2 + 2 u w c s + u^2 s^2) + w^2 s^2 - 2 u w c s + u^2 c^2
\beta^2(w^2 c^2 + 2 u w c s + u^2 s^2) + u^2 c^2 - 2 u w c s + w^2 s^2
\\
C_{1,2} & =
%
%-\beta^2 (v w c^2 + u v c s + t w c s + t u s^2) - v w s^2 + u v c s + t w c s - t u c^2
-\beta^2 (v w c^2 + (u v + t v) c s + t u s^2) - t u c^2 + (u v + t w) c s - v w s^2
\\
C_{2,2} & =
\beta^2 (v^2 c^2 + 2 t v c s + t^2 s^2) + t^2 c^2 - 2 t v c s + v^2 s^2
\end{align}
where $r_e$ is the galaxy effective radius in degrees,
$\beta$ is the axis ratio (\eqnref{eq:a}) and
we have defined the abbreviations
\begin{align}
\begin{bmatrix}
t & u \\
v & w
\end{bmatrix}
& = 
\begin{bmatrix}
\CD_{1,1} & \CD_{1,2} \\
\CD_{2,1} & \CD_{2,2}
\end{bmatrix}
=
\CD
\\
c &= \cos \theta \\
s &= \sin \theta
\end{align}
where the $\CD$ matrix terms have units of degrees per pixel.
Observe that the covariance terms only contain $c^2$, $cs$, and $s^2$
terms.  We can therefore use the identities:
\begin{align}
c^2 & = \cos^2 \theta            =
\tfrac{1}{2} \left(1 + \frac{e_1}{e} \right) \\
s^2 & = \sin^2 \theta            =
\tfrac{1}{2} \left(1 - \frac{e_1}{e} \right) \\
cs  & = \sin \theta \cos \theta  = %\tfrac{1}{2} \frac{e_2}{e}
\frac{e_2}{2 e}
\end{align}
where $e_1$ and $e_2$ are the two ellipticity components and $e =
\sqrt{e_1^2 + e_2^2}$.  This allows us to avoid computing $\theta$,
$\cos \theta$ and $\sin \theta$ explicitly when transforming a galaxy
into pixel space.


\begin{thebibliography}

\bibitem[Berg\'e \etal(2013)]{ufig}
Berg\'e,~J. \etal, 2013, Astronomy and Computing, 1, 23;
\arxiv{1209.1200}
% An Ultra Fast Image Generator (UFig) for wide-field astronomy


\bibitem[Bertin (2011)]{psfex}
Bertin, E., 2011, ADASS XX, ASP Conference Series, 442, 435
%Evans, I.~N., Accomazzi, A., Mink, D.~J. and Rots, A.~H., eds.

\bibitem[Calabretta \& Greisen(2002)]{wcs2}
Calabretta,~M.~R. \& Greisen,~E.~W., 2002, 
%\AA,
Astronomy \& Astrophysics, 
395, 1077;
\arxiv{astro-ph/0207413}

\bibitem[Dey \etal(2016)]{mosaic3}
  Dey,~A., Rabinowitz,~D. \etal, 2016,
  in Proc.~SPIE, 9908, Ground-based and Airborne
  Instrumentation for Astronomy VI, 99082C

\bibitem[Dey \etal(2019)]{dey}
  Dey,~A., Schlegel,~D.J., \etal, 2019,
  \aj, 157, 168;
  \arxiv{1804.08657}

\bibitem[Flaugher \etal(2015)]{decam}
  Flaugher,~B. \etal, 2015,
  \aj, 150, 150;
  \arxiv{1504.02900}
  
\bibitem[Hogg \& Lang(2013)]{moggalaxy}
Hogg,~D.~W. \& Lang,~D., 2013, \pasp, 125, 719;
\arxiv{1210.6563}

% Detailed Structural Decomposition of Galaxy Images
\bibitem[Peng \etal(2002)]{galfit}
 Peng,~C.~Y. \etal, 2002, \aj, 124, 266;
 \arxiv{0204182}

% PhoSim: a code to simulate telescopes one photon at a time
%Peterson,~J.~R., 2014, Journal of Instrumentation, 9, C04010;

%Simulation of Astronomical Images from Optical Survey Telescopes using a Comprehensive Photon Monte Carlo Approach
\bibitem[Peterson \etal(2015)]{phosim}
Peterson,~J.~R. \etal, 2015, \apjs, 218, 14;
\arxiv{1504.06570}

\bibitem[Rowe \etal(2015)]{galsim}
  Rowe,~B. \etal, 2015, Astronomy and
  Computing, 10, 121; \arxiv{1407.7676}
% GalSim: The modular galaxy image simulation toolkit

\bibitem[Sheldon (2014)]{ngmix}
  Sheldon,~E.~S., 2014, MNRAS, 444, 25;
  \arxiv{1403.7669}

\bibitem[Williams \etal(2004)]{90prime}
  Williams,~G.~G \etal, 2014,
  SPIE Conference Series, 5492, Ground-based Instrumentation for Astronomy, 787W

\end{thebibliography}
\end{document}